\documentclass[useAMS,usenatbib]{mn2e}

\usepackage{amsmath}
\usepackage{amssymb}
\usepackage[dvipsnames]{xcolor}
\usepackage{graphicx}
\usepackage{graphics}
\usepackage{hyperref}
\hypersetup{
     colorlinks   = true,
     citecolor    = blue
}

\def\apj{{ ApJ}}
\def\apjl{{ApJL}}
\def\apjs{{ ApJS}}

\def\aap{{ A\&A}}

\def\mnras{{ MNRAS}}

\def\nat {{ Nature}}

\def\physrep{{ Physics Reports}}

\def\nar{{New Astronomy Reviews}}

\def\mt{\mathrm}
\newcommand{\myemail}{wenbinlu@astro.as.utexas.edu}

\title[Dust IR Emission from TDEs]{Infrared emission from tidal
  disruption events --- \\ probing the pc-scale dust content around
  galactic nuclei} 
\author[Lu, Kumar \& Evans]
  {Wenbin Lu$^1$\thanks{\myemail},
  Pawan Kumar$^1$\thanks{pk@astro.as.utexas.edu},
  Neal J. Evans II$^1$\thanks{nje@astro.as.utexas.edu}\\
  $^1$Department of Astronomy, University of Texas at Austin, Austin,
TX 78712, USA}
\date{\today}

\pagerange{\pageref{firstpage}--\pageref{lastpage}} \pubyear{2015}
\def\LaTeX{L\kern-.36em\raise.3ex\hbox{a}\kern-.15em
    T\kern-.1667em\lower.7ex\hbox{E}\kern-.125emX}

\begin{document}
\label{firstpage}
\maketitle

\begin{abstract}
Recent UV-optical surveys have been successful in finding tidal
disruption events (TDEs), in which a star is tidally disrupted by a
supermassive black hole (BH). These TDEs release a  
huge amount of radiation energy $E_\mt{rad} \sim
10^{51} - 10^{52}\ \mt{erg}$ into the circum-nuclear medium. If
the medium is dusty, most of the radiation energy will be absorbed by
dust grains within $\sim 1\mt{\ pc}$ from the BH and re-radiated in
the infrared. We calculate the dust emission lightcurve from a 1-D 
radiative transfer model, taking into account the time-dependent
heating, cooling and sublimation of dust grains. We show that the dust
emission peaks at $3- 10\mt{\ \mu m}$ and has typical luminosities
between $10^{42}$ and $10^{43}\mt{\ erg\ 
  s^{-1}}$ (with sky covering factor of dusty clouds ranging from 0.1
to 1). This is detectable by current generation of telescopes. In the
near future, {\it James Webb} Space Telescope will be able to perform
photometric and spectroscopic measurements, in which silicate or
polycyclic aromatic hydrocarbon (PAH) 
features may be found.

Observations at rest-frame wavelength $\geq 2\rm \ \mu m$ have only
been reported from two TDE candidates, SDSS J0952+2143 and {\it Swift}
J1644+57. Although consistent with the dust emission from TDEs, the
mid-infrared fluxes of the two events may be from other
sources. Long-term monitoring is needed to draw a firm conclusion. We
also point out two nearby TDE 
candidates (ASSASN-14ae and -14li) where the dust emission may be currently
detectable. The dust infrared emission can give a snapshot of the 
pc-scale dust content around weakly- or non-active galactic nuclei,
which is hard to probe otherwise.
\end{abstract}


\section{Introduction}\label{sec:intro}
Most galaxies harbor weakly- or non-active central supermassive black
holes (BHs). Roughly once every $10^{4} - 10^{5}$ years in each
galaxy, a star enters the BH's tidal disruption radius within which
the tidal force of the BH exceeds the star's 
self gravity, and hence the star gets tidally disrupted
\citep{1982ApJ...262..120L, 1988Natur.333..523R,
  2004ApJ...600..149W}. In these so-called 
tidal disruption events (TDEs), the stellar debris feeds a burst of
strong accretion that generates a bright flare of 
electromagnetic radiation. Several dozen such flares have been
reported at various wavelengths,
e.g. hard X-ray/$\gamma$-ray \citep{2011Sci...333..199L,
  2011Sci...333..203B,   2012ApJ...753...77C}, soft X-ray
\citep[reviewed by][]{2015JHEAp...7..148K}, 
UV \citep[e.g.][]{2009ApJ...698.1367G, 2012Natur.485..217G} and
optical \citep[e.g.][]{2011ApJ...741...73V,
  2014ApJ...793...38A, 2015ApJ...798...12V, 2015arXiv150701598H}. 

If a large fraction of the stellar debris is accreted by the BH, the
total energy budget is $\sim \xi M_\mt{\odot} c^2\simeq  10^{53} \xi_{-1}\
\mt{erg}$, where $\xi=0.1 \xi_{-1}$ is the efficiency of energy
release from accretion and depends on the BH spin
\citep{1973A&A....24..337S}. However, 
the observed radiation energy is usually $10^{51} - 10^{52}\
\mt{erg}$, much smaller than expected. Especially, the energies of the
optically discovered TDEs lie at the lower end ($\sim10^{51}\
\mt{erg}$). A possible solution to the ``efficiency problem'' could be
that $\xi\ll0.1$ because most of the stellar debris doesn't reach the
BH but is blown away 
by a wind at a radius of $\sim10$ to 100 Schwarzschild radii
\citep{1997ApJ...489..573L, 1999ApJ...514..180U, 2009MNRAS.400.2070S,
  2015ApJ...805...83M, 2015arXiv150603453M, 2015Lu}. If the wind is
dense enough, the effective  
photosphere could be far from the surface of the disk, and the
radiation energy suffers from adiabatic loss before escaping, so the
spectrum appears redder and the total radiation energy lower. Another
possibility is that the spectral energy distritubion (SED)
peaks between the observable far
UV and soft X-ray windows.


We can see that TDEs release a huge amount of radiation energy, which
will have a significant impact on the interstellar medium in the
immediate vicinity of the BH. For example, as pointed out by
\citet{1999ApJ...514..180U}, if 
most of the flare's energy is radiated in the extreme UV ($\sim 10 -
100\ eV$), the bright flash can ionize gas of mass
\begin{equation}
  \label{eq:20}
  M_\mt{ion} \simeq \frac{1.3m_{H} E_\mt{rad}}{25\ \mt{eV}} \simeq 8.1\times
  10^4 M_\mt{\odot} E_{\mt{rad}, 51.5}
\end{equation}
out to a radius
\begin{equation}
  \label{eq:7}
  R_\mt{ion}\simeq \left( \frac{3}{4\pi n_\mt{H}}\frac{E_\mt{rad}}{25\ \mt{eV}}
  \right)^{1/3} \simeq 2.6 \times 10^{19} \left(\frac{E_{\mt{rad},
        51.5}}{n_\mt{H, 3}}\right)^{1/3} 
  \ \mt{cm}
\end{equation}
where $E_\mt{rad} = 3\times10^{51} E_{\mt{rad}, 51.5}\ \mt{erg}$,
$n_\mt{H} = 10^3n_\mt{H,3}\ \mt{cm^{-3}}$ is the number density of H
nuclei and we have assumed each 
H nucleus consumes $25\ \mt{eV}$ (ionization potential of H
and He + bond energy of H$_2$ + kinetic energy of free protons and
electrons). In eq.(\ref{eq:20}) and 
(\ref{eq:7}), recombination is ignored because the timescale 
$t_\mt{rec}\sim 10^3 n_\mt{H, 3}^{-1} T_4^{1/2}\ \mt{yr}$ is expected
to be long. Then the energy of the ionizing flash will be slowly
reprocessed into potentially observable H and He recombination lines
if the gas density is high, or forbidden lines (of metals) if the gas
density is low.

However, eq.(\ref{eq:20}) and (\ref{eq:7}) are true only if there is no 
dust extinction. If dust is well mixed with the gas, the V-band
extinction is proportional to the gas column density and we have
$A_\mt{V}/N_\mt{H}\simeq 5.3\times 10^{-22} \mt{\ mag\ cm^2\
    H^{-1}}$ for sightlines with $R_\mt{V} \equiv A_\mt{V}/E(B-V)=3.1$
  \citep{2011piim.book.....D}. 
The V-band extinction of the gas with a column depth of $N_\mt{H} =
n_\mt{H}R_\mt{ion} = 2.6\times10^{22} n_\mt{H,3}^{2/3} E_{\mt{rad},
  51.5}^{1/3}$ is
\begin{equation}
  \label{eq:22}
  A_\mt{V} \simeq 14 n_\mt{H,3}^{2/3} E_{\mt{rad},
  51.5}^{1/3}\mt{\ mag}
\end{equation}
Therefore, for our fiducial galactic center
density of $n_\mt{H} = 10^{3}\ \mt{cm^{-3}}$, the ionizing flash will be
significantly attenuated at a radius of $\sim2\times 10^{18}\
\mt{cm}$, unless dust is significantly depleted in the gas before the
TDE occurs.

The radiation energy will be deposited into dust grains,
which then re-radiate in the infrared (IR). A similar
process has been studied in the context of gamma-ray bursts (GRBs) by
\citet{2000ApJ...537..796W} and \citet{2002ApJ...569..780D}. As we
show in this work, if a significant amount of dust exists in the
pc-scale vicinity of the BH, the dust IR emission from TDEs is much
brighter than that from GRBs, lasts longer, and is hence easier to
observe.

Before moving forward to calculations of the dust IR emission, we
summarize our current (very limited) knowledge about the pc-scale dust 
content around galactic nuclei.

In the context of active galactic nuclei (AGN), the accretion disk
and broad line region are surrounded by a thick dusty torus
\citep[e.g.][]{2004Natur.429...47J, 2007A&A...474..837T,
  2008NewAR..52..274E}. The dusty torus absorbs a significant fraction
of nuclear luminosity and re-radiates mid-IR continuum as observed in
most AGNs. Typically, the sky covering factor of the 
torus is $\sim 1/2$ and inner boundary is roughly given by the dust sublimation
radius $\sim 0.4 L_{45}^{1/2}\ \mt{pc}$,
where $L = 10^{45} L_{45}\ \mt{erg}\ \mt{s^{-1}}$ is the disk
luminosity \citep{2008ApJ...685..160N}.  
The torus is dynamically active \citep{1988ApJ...329..702K}, possibly
coming from a clumpy wind 
launched from the accretion disk \citep{1992ApJ...385..460E,
  1994ApJ...434..446K}. For
weakly-active galactic nuclei with luminosities $L \lesssim 10^{42}\
\mt{erg}\ \mt{s^{-1}}$, the accretion disk can no longer sustain a
wind outflow rate required by the mid-IR emission extrapolated from
high-luminosity AGNs, so the torus is expected to disappear
\citep{2006ApJ...648L.101E}. This has not been observationally
confirmed, because the dust emission no longer dominates in the mid-IR
when $L \lesssim 10^{42}\ \mt{erg}\ \mt{s^{-1}}$
\citep{2015A&A...578A..74G}. As pointed out by
\citet{2006ApJ...648L.101E}, 
even at these luminosities, the disk outflow can still provide
a small but significant toroidal obscuration as long as the column density
exceeds $\sim 10^{21}\ \mt{cm^{-2}}$. 

In the context of non-active galactic nuclei, the nuclear dust content
is hard to probe due to the overwhelming starlight and telescopes'
limited angular resolution (1 arcsec $\sim$ 50 pc at a distance of 10
Mpc). Currently, the only observable case is our Galactic
Center. Radio observations (of tracer molecules) revealed   
$\sim 10^5 M_\odot$ of molecular clouds and diffuse molecular gas
at $\sim1 - 5\ \mt{pc}$ from the BH, mostly in a torus-like structure
(known as circum-nuclear disk, CND) with a sky covering factor $\sim
30\%$ viewed from the BH \citep[e.g.][]{2005ApJ...620..287H,
  2005ApJ...622..346C, 2012A&A...540A..50F}. Inward from the CND,  due
to the radiation 
from the central star cluster ($\sim 10^7 L_\odot$), the gas becomes
atomic and then ionized. The photo-dissociation region and HII
region are both dusty and are observable through atomic fine-structure lines
\citep{1985ApJ...293..445S, 1993ApJ...402..173J},  free-free continuum
\citep{1993ApJS...86..133R}, recombination lines
\citep{2003ApJ...594..294S}, and mid/far-IR dust thermal emission
\citep{2013ApJ...775...37L}. The densities of different regions
estimated by these works are $n_{\rm H}\sim10^3\rm\ cm^{-3}$ (HII
regions), $10^4\rm\ cm^{-3}$ (warm neutral medium and molecular
streamers), and $10^5\rm\ cm^{-3}$ (molecular cores).

The dust within a few parsecs of the Galactic Center may originate
from infalling molecular 
clouds/clumps \citep{1993ApJ...402..173J} or from in-{\it
  situ} dust formation \citep{2015Sci...348..413L}. Dust grains may 
be destroyed by two processes: sublimation and 
sputtering. The former may be unimportant because the dust sublimation
radius is $\ll 1\ \mt{pc}$ in weakly- or non-active galactic
nuclei. The latter process is significant only in the shock heated 
hot ionized medium where the gas temperature $\gtrsim 10^{5.5}\
\mt{K}$. 



We note the fact that the Galactic Center is unusually dusty may be
because the Milky Way is a gas-rich star-forming galaxy. The
nuclear dust content of elliptical galaxies is unknown but may be much less
than in our Galaxy. As we show in this work, the pc-scale dust content
around galactic nuclei can be probed by searching for the dust IR
emission after a TDE is discovered. In section \ref{sec:estimation}, we
give an order-of-magnitude estimation of the IR emission from the dust
heated by TDEs. In section \ref{sec:rt}, we calculate the bolometric lightcurve
from a 1-D time-dependent radiative transfer model. In section
\ref{sec:discussion}, we 
discuss the detectability and implications, as well as some possible
issues in our simple calculations. A short summary is given in section
\ref{sec:summary}. 

\section{Order-of-magnitude Estimation}\label{sec:estimation}
Lacking the knowledge of the dust distribution, we assume
spherical symmetry for simplicity. Consider a point source with
UV-optical luminosity $L(t)$ as a function of time $t$ in the rest
frame of the source. We 
assume a uniform spatial distribution of dust grains with number
density $n_\mt{d}$, spanning from inner radius $R_\mt{in}$ to outer
radius $R_\mt{out}$. As 
the UV-optical radiation propagates through the cloud, dust grains
will be heated up. For a grain of radius $a$ at radius $R$ from the
source, its temperature $T$ is determined by the equilibrium between
heating and radiative plus grain-sublimation cooling
\citep{2000ApJ...537..796W} 
\begin{equation}
  \label{eq:1}
  \mathrm{e}^{-\tau_\mt{UV}} \frac{L(t_r)}{4\pi R^2} \pi a^2 Q_\mathrm{UV} = \left<
    Q_\mathrm{abs} \right>_P 4\pi a^2 \sigma T^4 - 4\pi a^2
  \frac{\mathrm{d}a}{\mathrm{d}t} 
  \frac{\rho}{\mu} B
\end{equation}
where $\mt{e}^{-\tau_\mt{UV}}$ is the effective attenuation of the UV-optical flash
by inner dust shells, $t_r = t - R/c$ is the ``retarded'' time
(corresponding to the time segment of the UV-optical flash),
$Q_\mathrm{UV}\simeq 1$ 
\citep{2011piim.book.....D} is the absorption efficiency factor of
UV-optical radiation,
\begin{equation}
  \label{eq:2}
  \left< Q_\mathrm{abs} \right>_P \equiv \frac{\int B_\nu(T) Q_\mathrm{abs}(\nu)
    \mathrm{d}\nu}{\int B_{\nu}(T) \mathrm{d}\nu}
\end{equation}
is the Planck-averaged absorption efficiency factor, $\sigma$ is the
Stefan-Boltzmann constant, $\rho$ is the density of the grain
material, $\mu$ is the mean atomic mass, and $B$ is the chemical
binding energy. For the temperature range of interest in this work
$1000 \lesssim T \lesssim 2500\ K$, we use an intermediate absorption
coefficient between astronomical silicate and graphite
\citep{2000ApJ...537..796W}
\begin{equation}
  \label{eq:6}
  \left< Q_\mathrm{abs} \right>_P \simeq \frac{0.1 a_{-5} (T/2300\
    K)}{1 + 0.1 a_{-5} (T/ 2300\ \mt{K})}
\end{equation}
where the grain radius is normalized to a typical value of $a = 
10^{-5}a_{-5} \ \mathrm{cm}$. The thermal sublimation 
rate is a function of temperature and 
can be estimated by 
\begin{equation}
  \label{eq:3}
  \frac{\mathrm{d}a}{\mathrm{d}t} = - \nu_\mt{0} \left( \frac{\mu}{\rho}
  \right)^{1/3} \mathrm{e}^{-B/kT} 
\end{equation}
We take $\nu_\mt{0}\simeq 1\times 10^{15} \ \mt{s^{-1}}$, $B/k = 7\times 10^4 \
\mt{K}$, and $\rho/\mu = 1\times10^{23} \ \mt{cm^{-3}}$ as
representative values 
for refractory grains \citep{1989ApJ...345..230G,
  2000ApJ...537..796W}. From eq.(\ref{eq:3}), we get a
characteristic grain survival time at temperature T
\begin{equation}
  \label{eq:4}
  \begin{split}
     t_\mathrm{surv}(T) =& \frac{a}{|\mathrm{d}a/\mathrm{d}t|}  =
     4.9\times 10^6 a_{-5} 
     \\
& \cdot \mathrm{exp}\left[ 7\times10^4\ K\left( \frac{1}{T} -
    \frac{1}{1600\ \mt{K}} 
\right) \right] \ \mt{s}
  \end{split}
\end{equation}
If the duration of the TDE UV-optical flash is $t_\mt{TDE} = 10^6
t_\mt{TDE,6} \ s$, the ``sublimation temperature'' (above which dust
grains fully sublime) can be defined by 
$t_\mathrm{surv}(T_\mathrm{sub}) = t_\mt{TDE}$, i.e.
\begin{equation}
  \label{eq:5}
  T_\mathrm{sub} = 1.66\times 10^3 \left[ 1 + 2.37\times10^{-2} \mathrm{ln}
    \frac{t_\mt{TDE,6}}{a_{-5}} \right] \ \mt{K}
\end{equation}
It can be shown that the critical temperature at which radiative
cooling equals to sublimation cooling is $T_{r=s} \simeq 2800\
K$, so the sublimation cooling term in eq.(\ref{eq:1}) is usually negligible
and we can solve for the ``sublimation radius'' where the grain
temperature equals to $T_\mathrm{sub}$
\begin{equation}
  \label{eq:8}
  R_\mathrm{sub} = 1.4\times 10^{18} \left[L_{45.5} \mt{e}^{-\tau_\mt{UV}}
    a_{-5}^{-1} (1 + 0.072 a_{-5})   \right]^{1/2} \ \mt{cm}
\end{equation}
If the whole dusty layer is optically thick to the UV-optical radiation from
the central source, the total radiation energy $E_\mathrm{rad}= 3\times10^{51}
E_\mathrm{rad, 51.5}\ \mt{erg}$ will be absorbed and re-radiated at a
typical wavelength of $\lambda\simeq 3(T/1600\ \mt{K})^{-1}\ \mathrm{\mu 
m}$. Because the dust sublimation rate increases steeply
with grain temperature (eq. \ref{eq:3}), grains at radii
$R< R_\mt{sub}$ sublime fully before a significant fraction of UV-optical
energy is absorbed. From eq.(\ref{eq:5}), we can see that dust IR
emission usually cuts off at wavelength shorter than $3\mt{\ \mu m}$,
unless there is more dust at $R<R_\mt{sub}$ than the UV-optical
radiation can cause to sublime. The duration of the dust 
emission can be estimated by the 
light-crossing time $2R_\mathrm{sub}/c\simeq 6.5
R_\mt{sub, pc} \mt{\ yr}$, where $c$ 
is the speed of light and $R_\mt{sub, pc} = R_\mt{sub}/1\mt{\
  pc}$. The IR luminosity of the dust emission can be estimated by 
\begin{equation}
  \label{eq:9}
  \begin{split}
      L_\mathrm{IR} \simeq \frac{E_\mathrm{rad}
        f_\Omega}{2R_\mathrm{sub}/c} \mt{e}^{-\tau_\mt{IR}}
      \simeq 1.5\times 10^{43} \frac{E_\mathrm{rad,
          51.5}}{R_\mathrm{sub, pc}} f_\Omega \mt{e}^{-\tau_\mt{IR}} \
      \mt{erg/s}
  \end{split}
\end{equation}
where $f_\Omega$ is the sky covering factor of the dusty clouds viewed
from the BH and $\mt{e}^{-\tau_\mt{IR}}$ is the further attenuation of
the dust IR emission by outer dust shells that are not heated up. Note
that, when the dusty cloud is far enough 
from the central source that sublimation is not significant, most IR
emission comes from the inner edge of the cloud and $R_\mt{sub}$
should be replaced by $R_\mt{in}$. Also, eq.(\ref{eq:9}) is a conservative
estimate because when $f_\mt{\Omega} < 1$, say the plane of the dust
torus is oriented at an angle $\theta$ wrt. the line of sight, the
light-crossing time is a factor of $\cos\theta$ shorter than
$2R_\mt{sub}/c$. 
\section{Radiative Transfer}\label{sec:rt}
In this section, we model the heating and cooling of dust grains with
time-dependent radiative transfer in a 1-D spherical symmetric
grid. The bolometric lightcurve of dust IR emission is calculated.

\subsection{Model description}
Dust is distributed uniformly in the spherical layer
between the inner radius $R_\mt{in}$ and outer radius $R_\mt{out}$. We
assume all dust grains have the same initial radius $a_0$ and the
number density $n_\mt{d}$ is independent of radius. We divide the whole
dust layer into $N$ thin shells of thickness $\Delta R$ and the
middle point radius of the $j$-th shell ($j = 1, 2, 3, \ldots, N$) is
$R_j$. Due to sublimation, the grain radius in the $j$-th
shell $a_j(t_r)$ is a function of the retarded time $t_r$ (local time since
the arrival of the radiation front). The total optical depth from
$R_\mt{in}$ to $R_j$ is
\begin{equation}
  \label{eq:10}
  \tau_j = \sum_{i = 0}^{j-1} \Delta \tau_i
  = \sum_{i = 0}^{j-1} (\pi a_i^2 n_d Q_\mt{UV, ext} \Delta R)
\end{equation}
where the extinction coefficient factor $Q_\mt{UV, ext}= 2$ is from
nearly equal contributions from absorption 
and scattering at UV-optical wavelengths \citep{2011piim.book.....D}. Note
that the photons with scattering angle $\theta_\mt{s}$, after traveling a distance
$\delta R$, are delayed by
\begin{equation}
  \label{eq:11}
  \delta t = \frac{ \delta R(1 - \cos \theta_\mt{s})}{c} = 10^6
  \frac{\delta R}{10^{17}\ cm}
  \frac{1 - \cos\theta_\mt{s}}{0.3}\ \mt{s,}
\end{equation}
so the scattered photons suffer from significant delays. To keep the 
model simple and conservative, we 
ignore the absorption of the scattered photons. We also ignore the
small contribution to dust heating by the IR radiation
from other shells, because of the time delay and the extinction at IR
wavelength being small. For instance, $A_\mt{3\mu m}/A_\mt{V} =
0.069$ for $R_\mt{V} = 3.1$ \citep{1989ApJ...345..245C}. Actually, the
heating flux on dust  grains from UV and IR photons bouncing between
different layers is a factor of $\sim 10^2$ smaller than the flux
directly from the central source ($\sim E_\mt{rad}c/8\pi R^3$ instead of
$L/4\pi R^2$ on the left hand side of eq. \ref{eq:1}). Therefore, on
the light-crossing timescale (a few years), the grain
temperature is a factor of $\sim3$ smaller and hence
the reprocessed emission is around $10\rm \ \mu m$.

Combining eq.(\ref{eq:1}), (\ref{eq:2}) and (\ref{eq:3}), we use a
simple explicit scheme to solve the time evolution of grain
temperatures $T_j(t_\mt{r})$ and radii $a_j(t_\mt{r})$ in each
shell. We take the time step to be 
one tenth of the local grain survival time
\begin{equation}
  \label{eq:12}
  \Delta t = t_\mt{surv} (R)/10
\end{equation}
We have tested that the results are not affected by a factor of 2
variations in $\Delta t$ in all cases presented. As shown below, if we only
consider the dusty region to be at radius $R 
\gtrsim 0.3 \ \mathrm{pc}$, the grain temperatures are always smaller
than $T_{r=s}$ (the temperature at which the radiative cooling equals
to sublimation cooling). Since the sublimation cooling term in
eq.(\ref{eq:1}) drops fast with decreasing temperature [$\propto
\mathrm{exp}(-7\times10^4\ K/T)$], we ignore the sublimation cooling
term in our calculations.

From $\{T_j(t_r), a_j(t_r)\}$, we
get the volume emissivity ($\mt{erg} \ \mt{cm}^{-3}\
\mt{s}^{-1} \ \mt{sr}^{-1}$) of the $j$-th shell
\begin{equation}
  \label{eq:13}
  J_\mt{IR,j}(t_\mt{r}) = 4\pi a_j^2 n_d \left<
    Q_\mathrm{abs} \right>_P \sigma T_j^4/4\pi
\end{equation}
as a function of retarded time through $T_j(t_\mt{r})$ and 
$a_j(t_\mt{r})$. At any observer's time $t_\mt{obs}$, we can add up
the contributions from all 
the shells at different radii $R$ and emission latitudes $\theta$ to
get the total (isotropic equivalent) IR luminosity
\begin{equation}
  \label{eq:14}
  \begin{split}
      L_\mt{IR}(t_\mt{obs}) =& 4\pi \int_{R_\mt{in}}^{R_\mt{out}} R^2 \mt{d}R
  \int_0^\pi 2\pi \sin \theta \mt{d}\theta \\
&\cdot  J_{\mt{IR}} (R, t_\mt{r}) \mt{e}^{-\tau_{\mt{IR}} (R, \theta, t_\mt{r})}
  \end{split}
\end{equation}
where $e^{-\tau_\mt{IR}}$ is the attenuation of the IR emission from
position $(R, \theta)$ at $t_\mt{r}$ by other dust shells along
the line of sight. The observer's clock starts ($t_\mt{obs}=0$) when
the radiation front from 
the central source arrives, so the IR photons emitted from position
$(R, \theta)$ at retarded time $t_\mt{r}$ will reach the observer at
\begin{equation}
  \label{eq:15}
  t_\mt{obs} = R(1-\cos\theta)/c + t_\mt{r}
\end{equation}
which gives $t_\mt{r}$ as a function of $t_\mt{obs}$, $R$ and
$\theta$. 

To calculate the IR optical depth $\tau_\mt{IR} (R, \theta,
t_\mt{r})$, we integrate the absorption\footnote{Since the
  grain sizes are smaller than the typical dust emission wavelength
  ($\lambda \gtrsim 3\ \mt{\mu m}$), scattering has a smaller cross
  section than absorption, so we ignore the
  contribution to the IR optical depth from scattering.} by dust
shells along the line of sight.
\begin{figure}
  \centering
\includegraphics[width = 0.35 \textwidth,
  height=0.22\textheight]{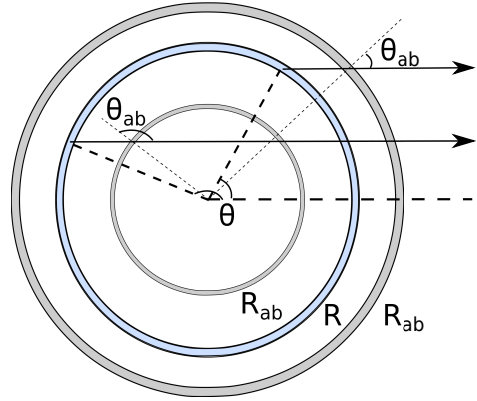}
\caption{The geometry of the extinction of the dust IR
  emission by the other shells along the line of sight. The observer is at
  infinity on the right-hand side. The IR light rays from the
  shell at radius R may intersect with inner or outer absorbing shells,
  whose radii are denoted as $R_\mt{ab}$. There are two possible
  cases: (1) if the emission latitude
  $\theta < 90^\mt{o}$, the line of sight intersects once with all the outer
  shells ($R_{ab} > R$); (2) if $\theta > 90^\mt{o}$, the line of sight
  intersects once with all the outer shells and twice with the inner
  shells at radii $R\sin\theta < R_\mt{ab} < R$.
}\label{fig:tauIR}
\end{figure}
As shown in Fig.(\ref{fig:tauIR}), there are two cases: (1) if
the emission latitude $\theta < 90^\mt{o}$, the line of sight
intersects once with all the outer shells ($R_\mt{ab} > R$); (2) if the $\theta >
90^\mt{o}$, the line of sight intersects once with all the outer shells and
twice with the inner shells at radii $R\sin\theta < R_\mt{ab} < R$. We 
denote the angle between the light ray and radial 
direction as $\theta_\mt{ab}$ and it satisfies
\begin{equation}
  \label{eq:16}
  \sin \theta_\mt{ab} = \frac{R \sin
        \theta}{R_\mt{ab}}
\end{equation}
To calculate the
optical depth contributed by each shell, we need to know the dust
content at the shell's retarded time when the IR light ray
arrives
\begin{equation}
  \label{eq:17}
  t_\mt{r, ab} = 
  \begin{cases}
    & t_\mt{r}  + \left[ R(1 - \cos \theta) \right]/c \mbox{, if
      $R_\mt{ab} > R$}\\ 
    & t_\mt{r}  + \left[ -R\cos \theta  + R_\mt{ab}\cos \theta_\mt{ab}
      \right] /c \mbox{, if $R_\mt{ab} < R$}
  \end{cases}
\end{equation}
Note that when the IR light ray arrives at an inner shell for the
first time, we have $\theta_\mt{ab} > 90^\mt{o}$ and hence
$\cos \theta_\mt{ab} < 0$. In the model, the grain number density
$n_d$ is fixed and the decreasing of dust content in a
shell (due to sublimation) is captured by the decreasing of grain
radii with its retarded time $a(t_\mt{r})$. When $a(t_\mt{r})$ is
smaller than $1/30$ of the initial grain 
radius $a_\mt{0}$, we consider the grain sublimation to be complete
and the shell 
becomes dust free. Therefore, we can 
calculate the optical depth of the IR emission from position $(R,
\theta)$ at $t_\mt{r}$: (1) the contribution from outer shells
\begin{equation}
  \label{eq:18}
  \tau_\mt{IR}^\mt{out} (R, \theta, t_\mt{r}) = 
    \int_R^{R_\mt{out}} \frac{dR_\mt{ab}}{|\cos
      \theta_\mt{ab}|} n_d \pi a^2 \left<
    Q_\mathrm{abs} \right>_P
\end{equation}
(2) the contribution from inner shells (only when $\theta > 90^\mt{o}$)
\begin{equation}
  \label{eq:19}
    \tau_\mt{IR}^\mt{in} (R, \theta, t_\mt{r}) = 
    \int_{R\sin \theta}^{R} \frac{dR_\mt{ab}}{|\cos
      \theta_\mt{ab}|} n_d \sum_{i = 1, 2} \left(\pi
      a_\mt{i}^2 \left<
    Q_\mathrm{abs} \right>_\mt{P,i} \right)
\end{equation}
where i = 1, 2 means the first/second time the IR light ray
intersects with the absorbing shell. Note that the Planck-averaged
absorption efficiency factor $\left< Q_\mathrm{abs} \right>_P$
(eq. \ref{eq:6}) depends on the grain size $a$ in the absorbing shell
(at $R_\mt{ab}$, $t_\mt{t, ab}$) and the radiation
temperature\footnote{The spectrum of dust emission from a thin shell
  at temperature $T$ is assumed to be Planckian. We note that, as the
  IR light ray propagates through the dust shells, the spectrum will
  change due to reddening. However, in all computed cases, the
  absorption at IR wavelength ($\lambda \gtrsim 3\ \mt{\mu m}$) is not
  very large,
  so the changing of spectrum is ignored.} $T$ of the emitting shell
(at $R$, $t_\mt{r}$).

\subsection{Results}
We run the 1-D radiative transfer model for four cases, which are summarized
in Table \ref{tab:cases}. The following parameters are fixed in all
four cases:
the radial thickness of the dusty cloud $R_\mt{out} - R_\mt{in} = 1\
\mt{pc}$, initial grain radius $a_0 = 0.1\ \mt{\mu m}$ (all grains
having the same initial radius) and the total dust optical depth
$\tau_{UV} = 10$ ($A_{V} \simeq 10\ \mt{mag}$). The
grain number density is constant with radius and 
is given by\footnote{If the density
of the grain material is $\rho = 2\ \mt{g}\ \mt{cm^{-3}}$, the grain
number density $n_\mt{d}$ corresponds to mass density of
$4.3\times10^{-23}\ \mt{g}\ \mt{cm^{-3}}$. Using a dust-to-gas mass
ratio of 0.01, we get a hydrogen number density of $n_\mt{H}\simeq
3\times 10^3 \ \mt{cm^{-3}}$.}
\begin{equation}
  \label{eq:21}
  n_\mt{d} = \frac{\tau_{UV}}{Q_\mt{UV,ext}\pi a_0^2 (R_\mt{out} -
    R_\mt{in})} \simeq 5.1\times10^{-9} \mt{cm^{-3}}
\end{equation}
where the extinction coefficient factor $Q_\mt{UV,ext} = 2$
(equal contributions from absorption and scattering).

\begin{table}
\centering
  \caption{Summary of the four cases, where we vary the inner edge of
    the dusty cloud $R_\mt{in}$, the UV-optical luminosity $L_0$ and duration
    $t_\mt{TDE}$ of the TDE source. The lightcurve of the source is
    assumed to be flat. The following parameters are fixed in all
    four cases: the radial
    thickness of the dusty cloud $R_\mt{out} - R_\mt{in} = 1\
    \mt{pc}$, initial grain radius $a_0 = 0.1\ \mt{\mu m}$ (all grains
    having the same initial radius) and the total dust optical depth
    $\tau_{UV} = 10$ ($A_{V} \simeq 10\ \mt{mag}$).} 
  \label{tab:cases}
  \begin{tabular}{@{}cccc@{}}
  \hline\hline
   Case No. &$R_\mt{in}$ [pc] & $L_0$ [erg/s] &
    $t_\mt{TDE}$ [s] \\
\hline
I & 0.3 & $3\times 10^{45}$ & $10^6$ \\
\hline
II & 1 & $3\times 10^{45}$ & $10^6$ \\
\hline
III & 1 & $1\times 10^{46}$ & $10^6$ \\
\hline
IV & 1 & $3\times 10^{44}$ & $10^7$ \\
\hline\hline
\end{tabular}
\end{table}

\begin{figure}
  \centering
\includegraphics[width = 0.45 \textwidth,
  height=0.25\textheight]{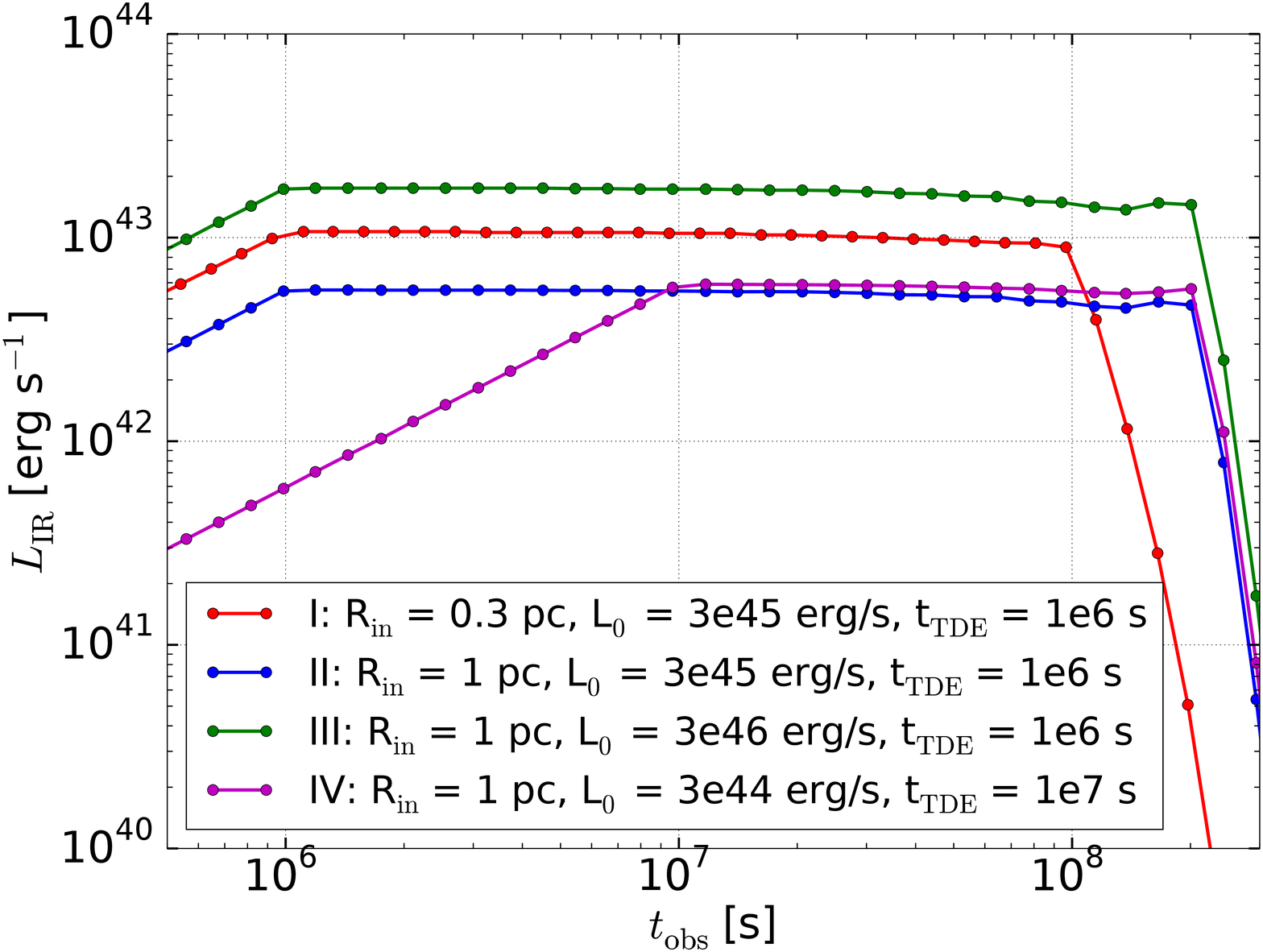}
\caption{Bolometric lightcurves of dust IR emission from the 1-D
  time-dependent radiative transfer model. The dust IR emission
  lasts for a few years with a typical luminosity of $\sim 10^{43}\
  \mt{erg} \ \mt{s^{-1}}$, consistent 
  with the estimation in eq.(\ref{eq:9}). The legend shows the variable
  parameters in each case and the fixed common parameters are
  described in the caption of Table \ref{tab:cases}.
}\label{fig:lc}
\end{figure}

\begin{figure}
  \centering
\includegraphics[width = 0.45 \textwidth,
  height=0.25\textheight]{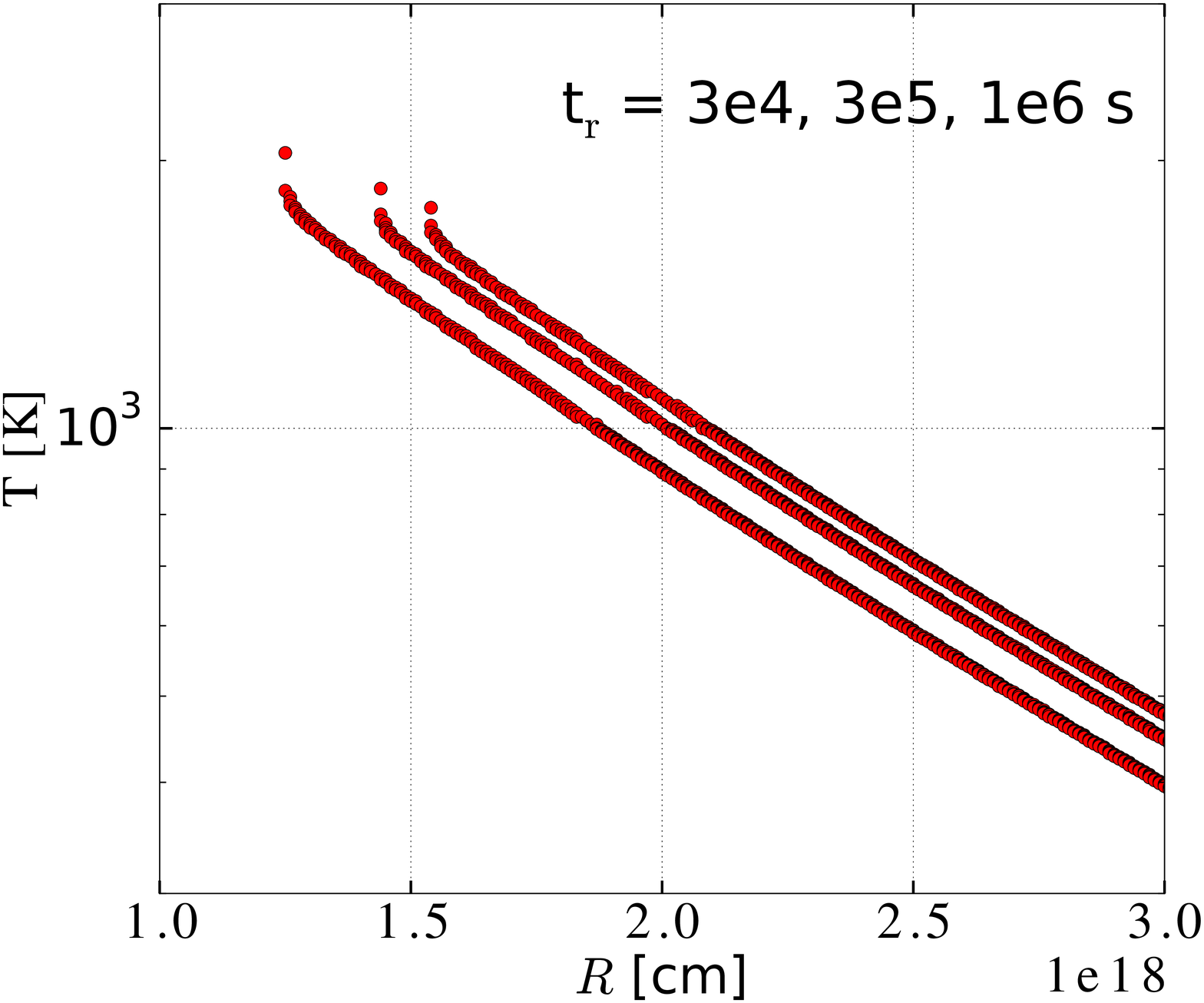}
\includegraphics[width = 0.45 \textwidth,
  height=0.25\textheight]{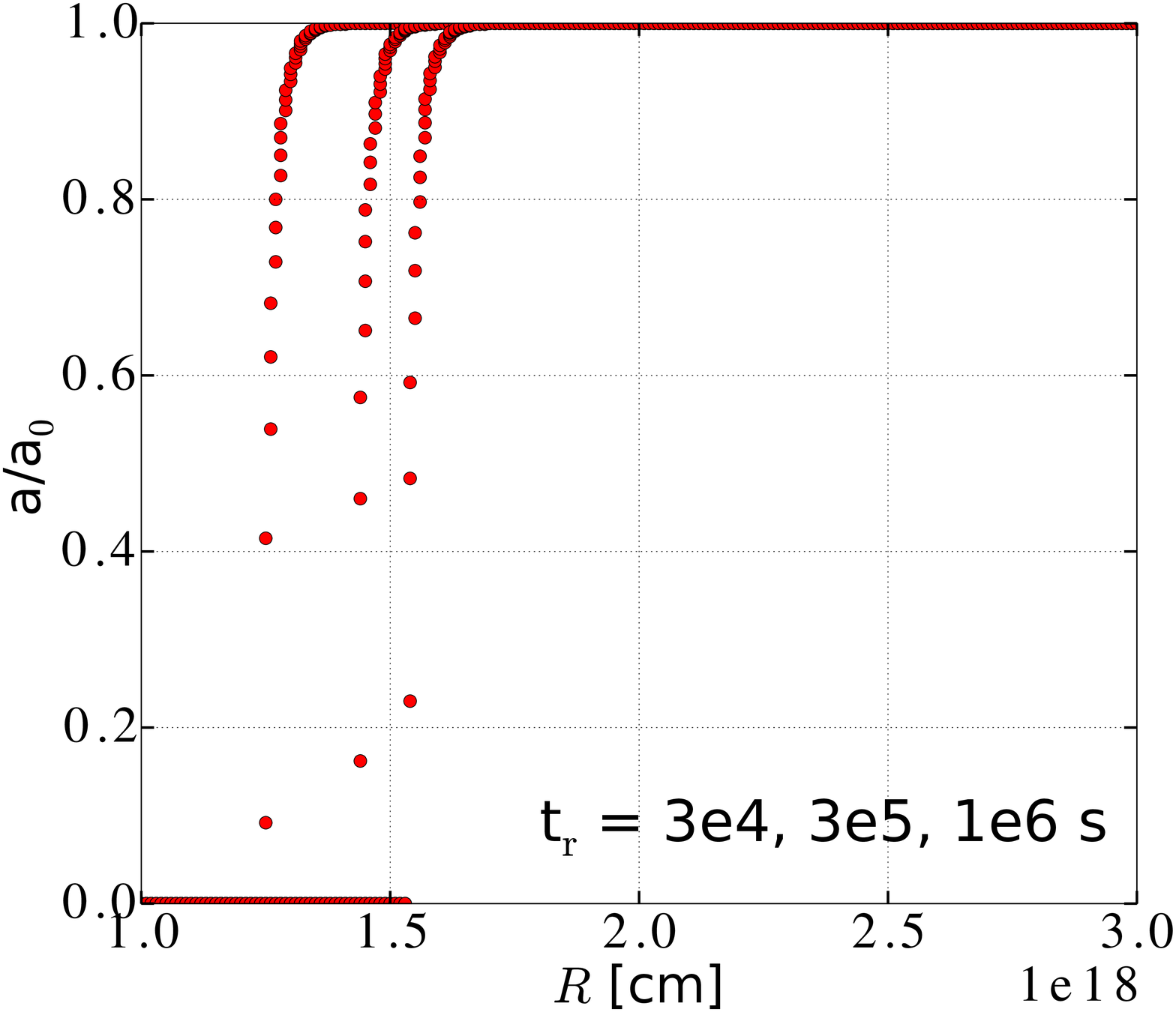}
\caption{The upper and low panels are respectively the grain
  temperature and radius profiles for case I (red curve  
  in Fig. \ref{fig:lc}) at different retarded time
 $t_\mt{r} = 3\times 10^4, 3\times 10^5, 1\times 10^6\ \mt{s}$. We can
 see that the dust sublimation front is very sharp and moves outwards
 with time. The temperature profile far beyond the sublimation front
 is exponential, as expected from the $\mt{e}^{-\tau_\mt{UV}}$ term in
 eq.(\ref{eq:1}). When $a(t_\mt{r})< a_0/30$, we consider the grain
 sublimation to be complete and set $a(t_\mt{r}) = 0$. This is the
 reason for the radius ``floor'' at small radii.
}\label{fig:profile}
\end{figure}

The bolometric lightcurves for all cases are shown in
Fig.(\ref{fig:lc}), from which we draw the following conclusions: (1)
the luminosities are determined by the total energy input $E_\mt{rad}$
and the light-travel delay, and are typically $\sim 10^{43}\ \mt{erg}
\ \mt{s^{-1}}$, consistent with the estimation in eq.(\ref{eq:9}); (2)
the lightcurves rise linearly until $t_\mt{TDE}$ (due to the increasing
emitting area) and when $t_\mt{obs} \gg t_\mt{TDE} $,
the lightcurves are nearly flat\footnote{The
small dip in each lightcurve towards the end corresponds to the
emission from altitude angle $\theta\simeq 90^\mt{o}$ (where the
projected surface area is slightly smaller).}. To illustrate the dust
sublimation process, in Fig.(\ref{fig:profile}) we show the grain
temperature and 
radius profiles for case I (red curve in Fig. \ref{fig:lc}) at different retarded time
$t_\mt{r} = 3\times 10^4, 3\times 10^5, 1\times 10^6\ \mt{s}$. We can 
see that the dust sublimation front is very sharp and moves outwards
with time. At $t_\mt{r} = t_\mt{TDE}= 1\times 10^6\ \mt{s}$, the dust
grains in the inner quarter (radius-wise) of the cloud have
fully sublimed. In case I, the IR
emission peaks at $\sim 3\ 
\mt{\mu m}$. The temperature profile far beyond the sublimation front
is exponential, as expected from the $\mt{e}^{-\tau_\mt{UV}}$ term in
eq.(\ref{eq:1}). We note that, at larger radii where the temperature
drops to much below 1000 K, the absorption coefficient adopted in
eq.(\ref{eq:6}) may not be a good approximation. However, since most
IR emission comes from the inner-most shells where the grain
temperatures are high, our results are not affected. 

\section{Discussion}\label{sec:discussion}
\subsection{Detectability}
As shown in section \ref{sec:estimation} and \ref{sec:rt}, for
$f_\Omega \sim 0.1 - 1$ (the fraction of UV-optical radiation from the
TDE intercepted by dust in the central few parsecs), the dust IR emission
has a typical bolometric luminosity of  
$L_\mt{bol}\simeq 10^{42} - 10^{43}\ \mt{erg\ s^{-1}}$ and peaks at
$\nu_\mt{peak}\simeq 9.4\times10^{13}  (T/1600\ \mt{K}) \ \mt{Hz}$
($\sim 3 \mt{\ \mu  m}$). For a luminosity distance of $d_\mt{L} = 1\
\mt{Gpc}$ (redshift $z\simeq 0.2$), the peak flux density is
$F_\mt{\nu, peak} \simeq L_\mt{bol}/(4\pi d_\mt{L}^2 \nu_\mt{peak}) 
\simeq 0.01 - 0.1\ \mt{mJy}$. This is photometrically detectable by current
generation of telescopes, including the Stratospheric
  Observatory for Infrared
  Astronomy\footnote{\href{https://www.sofia.usra.edu/}{https://www.sofia.usra.edu/}}
  (SOFIA, with the 
  FLITECAM and FORCAST instruments) and the {\it Spitzer} Space
Telescope\footnote{\href{http://www.spitzer.caltech.edu/}{http://www.spitzer.caltech.edu/}}
(hereafter {\it Spitzer}, with the IRAC 
instrument). In the near future, {\it James Webb} Space
  Telescope\footnote{\href{http://www.stsci.edu/jwst/science/sensitivity}{http://www.stsci.edu/jwst/science/sensitivity}}
  (JWST) 
will be able to perform photometric and spectroscopic
measurements.

We note that ground-based telescopes are much more  
sensitive at $2\mt{\ \mu m}$ (K band) than at $\gtrsim3\mt{\ \mu m}$.
Observations at $\gtrsim 3\ \mt{\mu m}$ by space telescopes are more
useful at detecting dust IR emission due to the following two reasons:
(1) at wavelengths close to the optical, the flux could be dominated
by other bright sources such as the accretion disk (as shown in
Fig. \ref{fig:SED} below) or star light; 
(2) since the extinction ratio $A_\mt{2\mu  m}/A_\mt{3\mu m} \simeq 2$
\citep{1989ApJ...345..245C}, dust emission at 
$\lesssim 2 \mt{\ \mu m}$ may be significantly absorbed by outer dusty
layers (and re-radiated at longer wavelengths). However, on one hand,
the confusion by the disk emission could be ruled out by long-term
monitoring, since the disk accretion rate is expected to drop fast
with time while the luminosity from dust emission stays more-or-less
constant for many years. On the other hand, a combination of high
UV-optical luminosities ($L_\mt{0}$) and small inner edge radii
($R_\mt{in}$) can lead to relatively high grain temperatures, such as in
case I (red curve in Fig. \ref{fig:lc}). For example, the flux density
ratio $F_\mt{2\mu m}/F_\mt{3\mu m}$ of a blackbody at $T = 1600\mt{\
  K}$ is only 0.72, so ground-based K-band observations may play an
important role in studying dust IR emission.

To the authors' knowledge, IR observations at rest-frame wavelength
$\geq2 \mt{\ \mu m}$ have only been reported from two TDE candidates,
SDSS J0952+2143  \citep{2008ApJ...678L..13K, 2009ApJ...701..105K} and
{\it Swift} J1644+57 
\citep{2015ApJ...808...96Y, 2015arXiv150908945L}. Below, we first
show that the mid-IR flux from SDSS J0952+2143, although consistent with
the dust emission from a TDE, is likely permanent emission
from dust-obscured starburst regions or an AGN. Then, we also show
that the mid-IR flux from {\it Swift} J1644+57 may
be produced by other radiation sources, such as the accretion disk or
the external shock driven by the interaction between disk outflow and
circum-nuclear medium. Therefore, we are not sure whether the dust
IR emission calculated in this work has been detected. At last, we
point out two TDE candidates 
where the dust IR emission may be currently detectable.

SDSS J0952+2143 (redshift z = 0.079) was reported as a TDE candidate
due to its strong but transient emission lines (of e.g. H, He, Fe),
which are interpreted as the light echo of a powerful EUV-X-ray
outburst (not directly observed). Photometry 
from Sloan Digital Sky Survey showed a fading optical continuum since
Dec 2004 and the outburst is
believed to have happened before this time. Near-IR (J, H and
K$_\mt{s}$-band) images showed quiescent (from 1998 to 2008) emission of
$L_\mt{NIR}\simeq 2.5\times 10^{43}\rm\ erg\ s^{-1}$ dominated by the
extended host galaxy. A {\it Spitzer}  mid-IR ($10 - 20\rm \ \mu
m$) spectrum obtained in Jun 2008 showed two pronounced silicate
peaks at $\sim10$ and $\sim 18\rm \ \mu m$, which 
resemble the SEDs of many Palomar-Green quasars
\citep{2007ApJ...666..806N}. The integrated luminosity was $L_{\rm
  10-20\mu m} = 3.5\times10^{43}\rm\ erg\ s^{-1}$. The total amount of
energy radiated in the observation window since the outburst is
$\sim 3.5\times 10^{51}\rm\ erg$ and the grain temperature can be
estimated to be $\sim 300-500 \rm\ K$. Therefore, the
mid-IR data is consistent with the scenario of emission from the
circum-nuclear dust heated by a TDE. However, as pointed out by
\citet{2009ApJ...701..105K}, the whole SED from optical to mid-IR
could also be produced by dusty starburst regions. In addition, an
X-ray luminosity of $10^{41}\rm \ erg\ s^{-1}$ may be due to a weak
AGN, which could also contribute some mid-IR flux. 
Another mid-IR observation is needed to tell whether the
mid-IR emission is transient or permanent.

{\it Swift} J1644+57 (redshift $z = 0.35$) is widely believed to be a
special TDE where a relativistic outflow is launched from the
accretion disk and produces 
bright emission from $\gamma$-ray to radio wavelengths
\citep[e.g.][]{2011Sci...333..199L, 2011Sci...333..203B,
  2011Natur.476..425Z}. This event received excellent observational
coverages in the time domain (a few to $\sim 1000$ days) and
wavelength domain ($\sim \mt{GHz}$ to $\sim\mt{TeV}$). Due to heavy
host extinction ($A_V\sim 10\ \mt{mag}$), the UV and optical fluxes are
strongly suppressed. For the purpose
of this work, we pay attention to the {\it Spitzer} $3.6\ \mt{\mu m}$
(rest-frame $\nu = 1.1\times10^{14}\ \mt{Hz}$) 
and $4.5\ \mt{\mu m}$ (rest-frame $\nu = 9.0\times10^{13}\ \mt{Hz}$)
data and see if it originates from the dust IR emission.

\begin{figure}
  \centering
\includegraphics[width = 0.45 \textwidth,
  height=0.25\textheight]{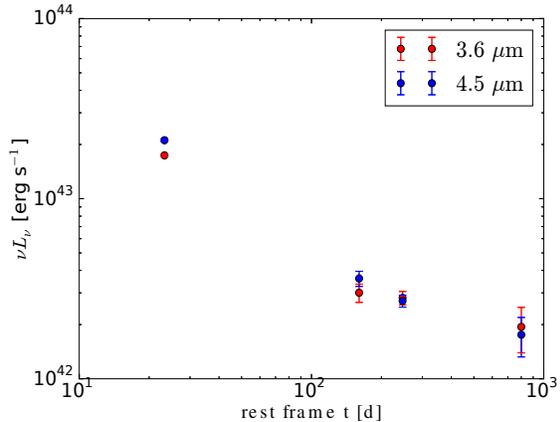}
\caption{Lightcurves of {\it Swift} J1644+57 at the {\it Spitzer} $3.6\
  \mt{\mu m}$ and $4.5\ \mt{\mu m}$ bands. The flux from the host
  galaxy is included and the extinction is not corrected.
  The data comes from
  \citet{2015ApJ...808...96Y, 2015arXiv150908945L}. 
}\label{fig:MIR_lc}
\end{figure}

In Fig.(\ref{fig:MIR_lc}), we plot the $3.6\ \mt{\mu m}$ and $4.5\
\mt{\mu m}$ band lightcurves, with the flux from the host galaxy
included and the host extinction is not corrected. The latest data
points (at rest-frame $t\sim 800\ \mt{d}$) are considered as host
flux, although the lightcurves are not completely flat. In a similar
way, the host flux in J, H, and K bands are obtained and reported by
\citet{2015arXiv150908945L}. In Fig.(\ref{fig:SED}), we plot 
the SEDs with the radio data from
\citet{2013ApJ...767..152Z} and the host-subtracted IR data from
\citet{2015ApJ...808...96Y, 2015arXiv150908945L} at rest-frame time
$t\simeq 150\mt{\ d}$ and $250\mt{\ d}$.

On the high-frequency side, the $3.6\ \mt{\mu m}$ and $4.5\
\mt{\mu m}$ fluxes are consistent with being the Rayleigh-Jeans tail
of a blackbody component (especially in the lower panel at rest-frame
$t \simeq 250\mt{\ d}$). As pointed out by
\citet{2015arXiv150908945L}, the blackbody component may originate
from the accretion disk.

On the low-frequency side, the radio emission has been extensively
studied and is well explained by 
the external shock produced by the interaction between the disk
outflow and circum-nuclear medium \citep{2012MNRAS.420.3528M,
  2013ApJ...770..146B,  
  2013ApJ...767..152Z, 2013MNRAS.434.3078K, 2015MNRAS.450.2824M}. The
details of the outflow are different in different models, but at
$\gtrsim 200$ days, the blast wave should more or less 
settle to the Sedov-Taylor evolution. In the standard external shock
model \citep[see e.g.][]{2015PhR...561....1K}, electrons are
accelerated to a relativistic powerlaw distribution and then produce a
broken powerlaw spectrum via synchrotron radiation. From
Fig.(\ref{fig:SED}), we can see that, if 
there is no spectral break between the radio and IR, the $3.6\ \mt{\mu m}$
and $4.5\ \mt{\mu m}$ fluxes are consistent with the extrapolation of
the powerlaw around $10^{11}\mt{\ Hz}$. For an electron
energy distribution of $dN/d\gamma_e\propto \gamma_e^{-p}$, a
spectral break may be produced at the synchrotron cooling frequency
$\nu_c$, because $\nu F_\nu\propto \nu^{(1-p)/2}$ when $\nu<\nu_c$ and
$\nu  F_\nu\propto \nu^{1-p/2}$ when $\nu>\nu_c$. If the blast wave is
sub-relativistic, the synchrotron cooling
frequency is given by  
\begin{equation}
  \label{eq:24}
  \nu_c = \frac{27\pi qm_ec}{\sigma_T^2B^3t^2} \simeq 8.4\times 10^{15}
  \left( \frac{B}{10\mt{\ mG}}\right)^{-3} \left(\frac{t}{200\mt{\
        d}}\right)^{-2} \mt{\ Hz} 
\end{equation}
where $m_e$ is electron mass, $q$ is electron charge, $\sigma_T$ is
Thomson scattering cross-section, $B$ is the strength of magnetic
field, $t$ is the dynamical time. From eq.(\ref{eq:24}), we can see
that, as long as $B\lesssim 50\mt{\ mG}$, the synchrotron cooling
break is expected to be above $10^{14}\mt{\ Hz}$, and hence the
$3.6\ \mt{\mu m}$ and $4.5\ \mt{\mu m}$ fluxes are consistent with
being produced by the external shock. The magnetic field energy density
in the shocked region is usually assumed to be a fraction $\epsilon_B$
of the thermal energy density and hence depends on the circum-nuclear
medium density and $\epsilon_B$. These two parameters are very
uncertain, so a firm conclusion cannot be drawn.
\begin{figure}
  \centering
\includegraphics[width = 0.45 \textwidth,
  height=0.25\textheight]{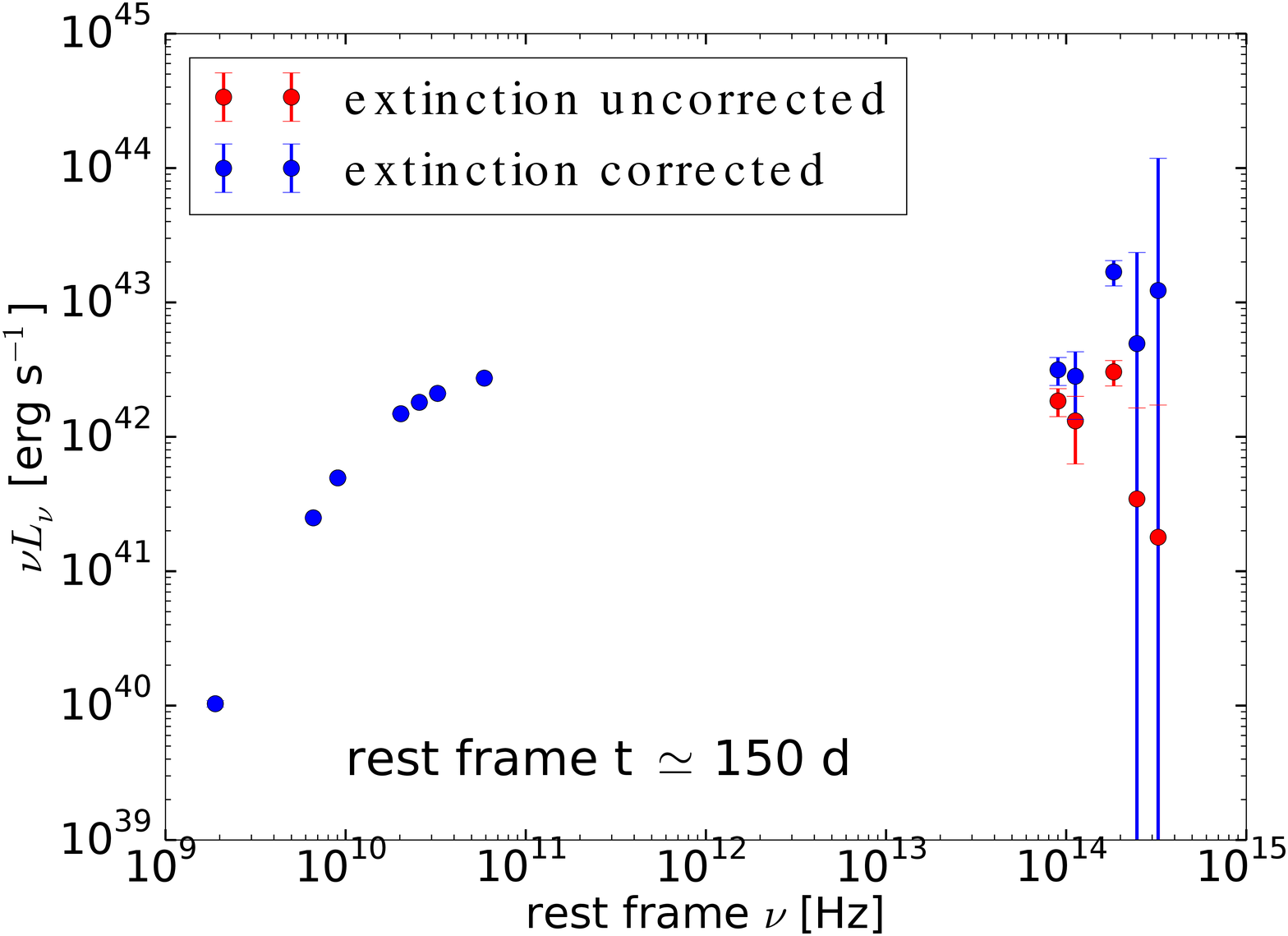}
\includegraphics[width = 0.45 \textwidth,
  height=0.25\textheight]{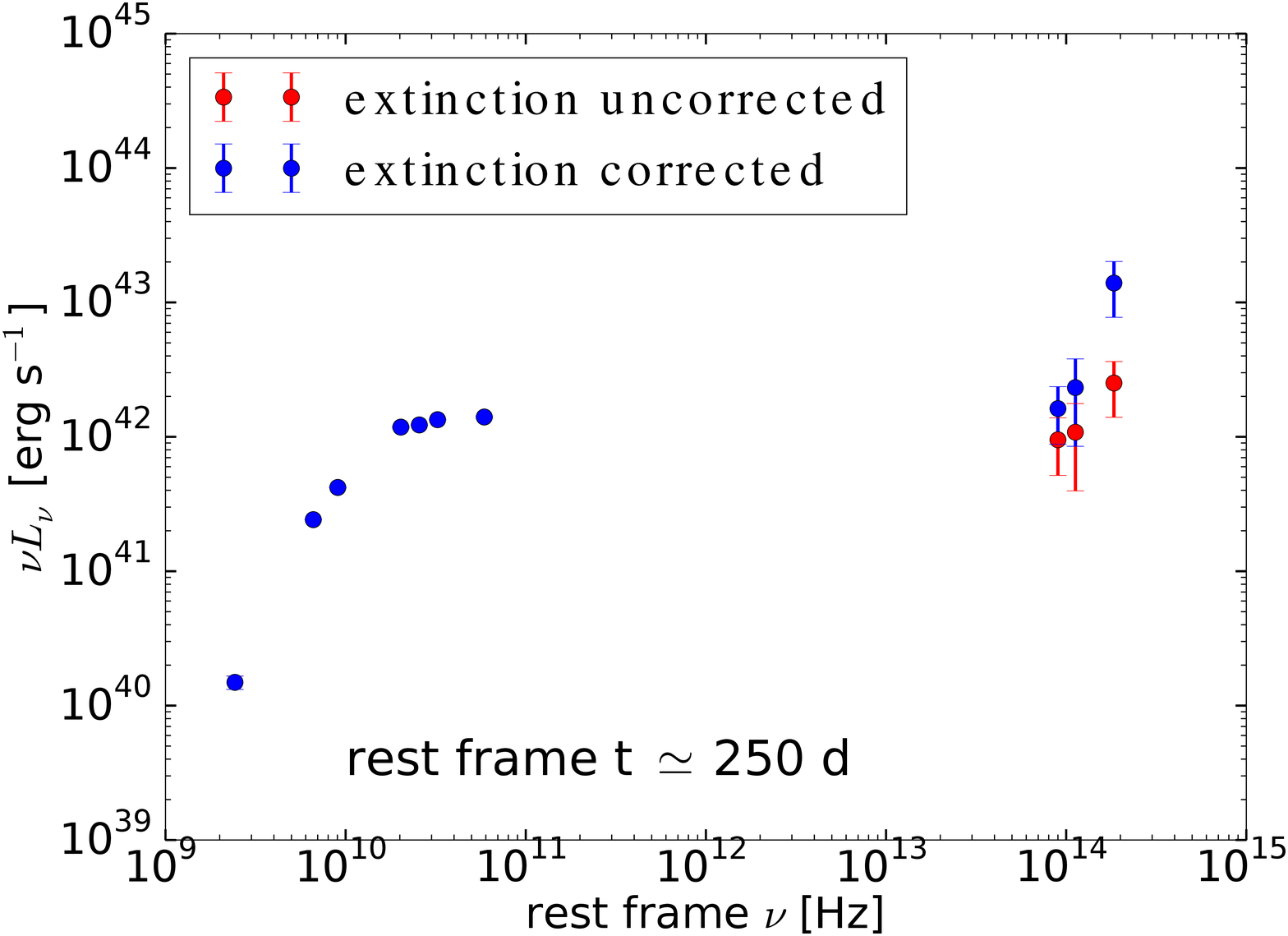}
\caption{Spectral energy distributions of {\it Swift} J1644+57 at
  rest-frame time $t\simeq 150\mt{\ d}$ (upper panel) and $250\mt{\
    d}$ 
  (lower panel). The radio data comes from
  \citet{2013ApJ...767..152Z} and the radio flux from the host galaxy
  is negligible. The IR data comes from \citet{2015ApJ...808...96Y,
    2015arXiv150908945L}. The red points are host-subtracted
  fluxes. The blue points are further corrected for an extinction of $A_V =
  10\mt{\ mag}$, following \citet{1989ApJ...345..245C} with $R_{V} = 3.1$.
}\label{fig:SED}
\end{figure}

On the other hand, since the dust IR emission typically lasts for a
few years, we encourage observations at $3-10\mt{\ \mu m}$ on some
recently-discovered TDE candidates, especially the most nearby ones
ASSASN-14ae \citep{2014MNRAS.445.3263H} and -14li
\citep{2015arXiv150701598H}. The reported luminosities and total
radiation energies above are respectively
$L\simeq 8\times 10^{43}\mt{\ erg\ s^{-1}}$ and  
$E_\mt{rad} \simeq 2\times 10^{50}\mt{\ erg}$ (former),
$10^{44}\mt{\ erg\ s^{-1}}$ and $7\times10^{50}\mt{\ erg}$
(latter). We note that, if the SEDs of these two events are 
reddened by host-galaxy extinction, the true luminosities and energies
are higher. Taking the reported $L$ and $E_\mt{rad}$ as conservative
values and assuming the pc-scale dust covering factor $f_\mt{\Omega} =
0.1$, we expect the peak flux densities to be 
$\sim 0.03\mt{\ mJy}$ for ASSASN-14ae and $\sim 0.5\mt{\ mJy}$ for
ASSASN-14li and peak wavelengths to be between 3 and 10 $\mt{\mu m}$
for both events.

To confirm the nature of dust IR emission, other possibilities can be
ruled out in the following way: (1) unlike {\it Swift} J1644+57,
the two ASSASN TDE candidates are not expected to have strong
relativistic outflows and are hence radio quiet. Actually,
\citet{2015arXiv151001226A} recently reported weak radio emission from
ASSASN-14li, which was interpreted as a weak outflow driving an
external shock. However, extrapolation from the radio to IR
gives a flux much smaller than the predicted dust
emission, so there is no contamination from the external shock. (2)
With a good multi-wavelength coverage from mid-IR to 
optical, other components (such as the disk emission) could be
separated out. (3) The contribution from the host galaxies could be
subtracted by multi-epoch or high spacial resolution observations.


In future JWST mid-IR spectra, we can look for the smoking-gun
evidences of dust, such as the silicate feature at $9.7\mt{\ \mu m}$
or polycyclic aromatic hydrocarbon (PAH) features.

\subsection{Other observable signatures?}
(1) If a significant amount of dust exists within $\sim10^{18}\mt{\
  cm}$, the 
refractory elements Al, Ca, Si and Fe, which normally condense into
solids, will gradually return to the gas phase. Observing their
abundances rising with time would be a direct evidence of dust
sublimation. The abundances could be measured through absorption
lines. 

(2) If the dusty clouds lie along the line of sight, as dust grains
sublime, the time evolution of dust extinction may be observable with
sufficient UV and optical wavelength coverages.

(3) For simplicity, we have been assuming the source lightcurve to be
flat, but TDEs actually have UV and optical variability down
to at least timescales of days. Therefore, it may be possible to
measure the time dependent dust IR emission in
response to the variable UV and optical continuum. This so-called
``dust reverberation mapping'' method has been successfully used to
determine the sublimation radius of the dusty torus in AGNs since the
work by \citet{1989ApJ...337..236C}.
\subsection{Some potential issues}
In this subsection, we discuss some potential issues that may affect
the details of the dust IR emission (but the general result in
this work is unaffected).

(1) To solve the grain temperature, we have assumed instantaneous
thermal equilibrium in eq.(\ref{eq:1}). This is valid because the
heat capacity of a single grain is small, and hence the cooling time
is short. Regardless of the composition, the upper limit of the
total internal energy of a grain is $(\rho/\mu)(4\pi/3)a^3B\simeq
4\times 10^{-3} a_{-5}^3\mt{\ erg}$. At a temperature of $T = 10^3
T_3\mt{\ K}$, the radiative cooling rate is $\left<Q_\mathrm{abs}
\right>_P 4\pi a^2 \sigma T^4 \simeq 3\times 10^{-3} a_{-5}^3 T_3^5
\mt{\ erg\ s^{-1}}$. Therefore, the cooling time is $< 1
T_3^{-5}\mt{\ s}$, which means that dust grains reach thermal
equilibrium nearly instantaneously.

(2) We have ignored dust destruction processes other than thermal
sublimation. As mentioned by \citet{2000ApJ...537..796W} in the
GRB context, photo-ionization of dust grains 
could lead to significant grain charging (positive). The electric
field at the surface can be so large that the grain fragments (``Coulomb
explosion'') or emits individual ions (``ion field emission''),
depending on how strong the grain structure is. If the grain charge is
$Z_\mt{g} = 10^4 Z_\mt{g,4}$ (in electron units), the electrostatic
binding energy of an electron at the grain surface is $Z_\mt{g}q^2/a\simeq
140 Z_\mt{g,4}a_{-5}^{-1}\mt{\ eV}$ ($q$ being the electron
charge). Photo-electric charging stops when the electrostatic binding
energy exceeds the incoming photon's energy. When exposed
to the intense X-ray radiation from
GRBs, the grain charge reaches $Z_\mt{g}\simeq 10^{5}$
\citep{2001ApJ...563..597F}; but in the 
context of TDEs where the
radiation energy is in UV and optical, the grain charge $Z_\mt{g}$ is
smaller than $10^4$. The
surface electric field is $Z_\mt{g}q/a^2 = 1.4\times10^7
Z_\mt{g,4}a_{-5}^{-2}\mt{\ V\ cm^{-1}}$, corresponding to a
stress $S = E^2/4\pi\simeq 1.8\times 10^8 Z_\mt{g,4}^2
a_{-5}^{-4}\mt{\ dyne\ cm^{-2}}$. Ion field emission
happens when the electric field approaches $\sim 10^{8} \mt{\ V\
  cm^{-1}}$ (to break a typical $1\mt{\ eV}$ molecular bond across a
distance of $1\mt{\ \AA}$). Coulomb explosion happens when the
electrostatic stress is greater than the tensile strength of grains,
which is uncertain. In the GRB
literature \citep[e.g.][]{2001ApJ...563..597F}, a typical grain
tensile strength $S_\mt{max}
\simeq 10^{10} \mt{\ dyne\ cm^{-2}}$ is used. Therefore, once the
grain charge reaches $Z_\mt{g}\simeq 10^{5}$, both Coulomb
explosion and ion field emission can happen. However, neither of these
two grain destruction mechanisms are important in the TDE context
because the grain charge is too small. Even for relativistic TDEs like {\it Swift}
J1644+57 with intense X-ray emission, since the X-rays are
relativistically beamed into a narrow angle, most dusty clouds only
receive the more-or-less isotropic UV-optical radiation.

(3) A few other complexities, which are ignored in our simple model, can be
taken into account in future works. First, the heating by the
scattered UV-optical photons and dust IR
radiation could be included by considering the time delay and related
absorption coefficient. In this way, a spectrum of dust IR emission can be
calculated and compared with future observations. Second, the
clumpiness of dusty clouds will affect the radiative transfer. Since
every clump will have a bright (hot) side and a dark (cold) side, dust
temperature distributions are different in clumpy and smooth
environments. This is considered in modeling the IR emission from the
AGN torus \citep{2008ApJ...685..160N}. Last, in the radiative
transfer model in section \ref{sec:rt}, a single dust radius of $a =
0.1\mt{\ \mu m}$ is used, but the true interstellar grain population
has a broad size distribution, extending from a few $\mt{\AA}$ (PAH
molecules) to about $0.2\mt{\ \mu m}$ (or even larger). Most mass is
in the larger grains,
with half-mass grain radius typically $\sim0.1\mt{\ \mu m}$ (half of
mass is in grains with radius above this value), and the smaller
grains contribute most surface area and hence most UV extinction
\citep[e.g.][]{2001ApJ...548..296W, 1993ApJ...402..441L}. However,
at the same distance from the central source, smaller grains sublime
faster than larger ones (mostly due to higher grain temperatures), so it
is unclear which grain size contributes most IR emission. A systematic
study of grain size distribution and chemical composition could be done in
the future with dust continuum radiative transfer codes such as
DUSTY\footnote{\href{http://www.pa.uky.edu/~moshe/dusty/}{http://www.pa.uky.edu/$\sim$moshe/dusty/}} and
HYPERION\footnote{\href{http://www.hyperion-rt.org/}{http://www.hyperion-rt.org/}},
although the dynamical 
propagation of radiation coupled with dust sublimation requires
some modification of the codes.
\section{Summary}\label{sec:summary}
We consider the impact of the UV-optical radiation ($E_\mt{rad}\sim
10^{51} - 10^{52}\mt{\ erg}$) from TDEs on the circum-nuclear
medium. If the medium is dusty, the UV-optical radiation energy will
be absorbed by dust within $\sim1\mt{\ pc}$ from the BH and
re-radiated in the IR. We calculate the bolometric lightcurve from a 1-D 
radiative transfer model, taking into account the time-dependent
heating, cooling and sublimation of dust grains. We show that the dust
emission  peaks at $3- 10\mt{\ \mu m}$ and has typical luminosities
between $10^{42}$ and $10^{43}\mt{\ erg\ 
  s^{-1}}$ (with sky covering factor $f_\mt{\Omega}$ ranging from 0.1
to 1). This is detectable by current generation of telescopes,
including the Stratospheric Observatory for Infrared Astronomy (SOFIA)
and the {\it Spitzer} Space Telescope. In the near future, {\it James
  Webb} Space 
Telescope (JWST) will be able to perform photometric and spectroscopic
measurements. In future JWST mid-IR spectra, we can look for the
smoking-gun evidence of dust, such as the silicate or
polycyclic aromatic hydrocarbon (PAH) features. 

Observations at rest-frame wavelength $\geq 2\rm\ \mu m$ have only
been reported from two TDE candidates, SDSS J0952+2143
\citep{2008ApJ...678L..13K, 2009ApJ...701..105K} and {\it Swift}
J1644+57  \citep{2015ApJ...808...96Y, 2015arXiv150908945L}. The {\it
  Spitzer} $10-20\rm \ \mu m$ flux of SDSS 
J0952+2143, although consistent with the dust emission from a TDE, is
likely permanent 
emission from dusty starburst regions or an AGN. We also show that the
{\it Spitzer} $3.6\ \mt{\mu m}$ and $4.5\ 
\mt{\mu m}$ fluxes of {\it Swift} J1644+57 may originate from
other sources than the dust IR emission, such as the accretion disk or
the external shock (driven by the interaction between disk outflow and
circum-nuclear medium). Long-term monitoring is
need to tell whether the mid-IR emission from the two event is
transient or permanent. Currently, a firm conclusion cannot be drawn.
At last, we show that the two nearby TDE candidates ASSASN-14ae
and -14li are good candidates to search for the dust IR emission. If
the pc-scale dusty clouds have a sky covering factor of $f_\mt{\Omega} =
0.1$, the conservatively estimated fluxes for the two events are,
respectively, $0.03\mt{\ mJy}$ and $0.5\mt{\ mJy}$ at $3- 10\mt{\ \mu
  m}$.

AGNs are known to be surrounded by a dusty torus that provides
obscuration and IR emission. However, little is known about the dust
content in the pc-scale vicinity of weakly- or non-active galactic
nuclei. Currently, the only observable example is our Galactic Center,
where the dusty clouds has sky covering factor $f_\mt{\Omega} \sim
30\%$ \citep{2012A&A...540A..50F}. The dust IR emission can give a
snapshot of the pc-scale dust content 
around weak- or non-active galactic nuclei, which is hard to probe
otherwise.
\section{acknowledgments}
We acknowledge helpful discussions on the Galactic Center environment
with J. Lacy. This research was funded by a graduate fellowship  (named
``Continuing Fellowship'') at the University of Texas at
Austin. N.J.E. acknowledges support from a grant from the National
Science Foundation, AST-1109116.

\label{lastpage}
\end{document}